\documentclass[aps,prd,reprint,superscriptaddress,showpacs,showkeys,nofootinbib]{revtex4-1}
\usepackage{pstricks,epsfig,color}
\usepackage{dcolumn}
\usepackage{bm}
\usepackage{float}
\usepackage{url}
\begin{document}

\title{Scalar Leptons in Folded Supersymmetry}

\author{Gustavo Burdman}
\affiliation{School of Natural Sciences, Institute for Advanced Study, Princeton, NJ 08540, USA}
\affiliation{Instituto de F\'{i}sica, Universidade de S\~{a}o Paulo,
  S\~{a}o Paulo SP 05508-900, Brazil~\footnote{Permanent address.}}
\author{Raffaele Tito D'Agnolo} 
\affiliation{School of Natural Sciences, Institute for Advanced Study, Princeton, NJ 08540, USA}


\begin{abstract}
Folded supersymmetry is a natural theory of the electroweak scale in
which the  scalar top partner responsible for canceling the
ultraviolet sensitivity of the Higgs mass at one loop carries no
color. As a result, bounds on naturalness on these theories are more
relaxed than in typical supersymmetric models, since collider bounds
on top partners are less stringent. Here we consider the lepton sector
of these theories. We show that a natural realization in a
five-dimensional completion requires the
existence of scalar leptons with weak scale masses and study their
phenomenology. We derive  bounds on the masses of these
sleptons  from the LHC data at $\sqrt{s}=8~$TeV and study the
sensitivity of the $\sqrt{s}=13~$TeV run for various integrated
luminosities. When supersymmetry is broken exclusively by
Scherk-Schwarz boundary conditions the sleptons are stable on collider
scales. This leads to bounds on the folded sleptons that 
translate into masses above  $1~$TeV for the folded stops resulting in
tuning. When additional sources of supersymmetry breaking, which may be necessary to ensure electroweak symmetry
breaking, are allowed the bounds on stops are mostly avoided. On the other
hand, these terms lead to  faster slepton decays resulting in interesting signals with highly displaced
vertices and multi-jet final states inside the detectors. 

\end{abstract}



\maketitle

\section{Introduction}
\label{sec:intro}
With the recent discovery of the Higgs boson at the
LHC~\cite{Aad:2012tfa,Chatrchyan:2012ufa} the standard model (SM)
spectrum is now complete. Furthermore, to the extent that they have been
tested, the Higgs boson couplings appear to be in agreement with the
SM predictions. When we add this new information from the Higgs sector
of the theory to the previously existing measurements on the gauge
interactions, one can conclude that the SM not only appears to be valid 
at the weak scale, but is probably valid at energies well above
it. Among the reasons to conclude that this extrapolation is warranted
are the apparent consistency of the Higgs potential with only renormalizable
interactions and the lack of deviations of the Higgs couplings to SM
gauge bosons and fermions. Finally, there is the absence of new states
at the weak scale. 

Although it might be too early in the life of the
LHC to assume that this situation will stand, it is interesting to
examine the consequences it would have on naturalness. As it is and
with $m_h=125~$GeV, if it
is valid up to energies well above the weak scale the SM is fine
tuned, with the tuning roughly given by $v^2/M^2$, with $v\simeq
246~$GeV and $M$ the SM ultraviolet (UV) scale. The absence of
deviations from the SM, both in electroweak precision measurements and
from direct searches, seem to point to $M\agt$~few TeV.
Models that try to explain this gap between the scales $v$ and $M$,
the so-called little hierarchy, need to protect $m_h$ in one of two
ways: through a spontaneously broken global symmetry, or
supersymmetry. In the first case, the Higgs is a
pseudo-Nambu-Goldstone boson (pNGB) and the theory must be
UV-completed above $M$ by a new strong dynamics. In general, the Higgs
couplings are modified and new states are present around the weak
scale, some of which carry color. 
On the other hand, supersymmetric
extensions of the SM do not significantly modify Higgs couplings,
although they predict several new states not far from the weak
scale, including color-carrying scalars.

These models become less natural as the bounds on new states become
stronger.  These constraints are mostly
driven by the LHC searches of new colored states. In the case of
composite Higgs models, the precise measurements of the Higgs 
couplings are  an additional constraint. This situation is somewhat
alleviated in models that realized the so-called neutral
naturalness. These are theories where the symmetry partners of the top
quark responsible for the cancellation of the top contribution to the
Higgs mass UV sensitivity, do not carry $SU(3)_c$ quantum numbers. 
When the Higgs is protected by a spontaneously broken global symmetry
(pNGB Higgs) these are
called Twin Higgs models~\cite{Twnhggs2006}. The protection is done
by supersymmetry in the so-called Folded Supersymmetric
models~\cite{fldsusy2007}, which are the  focus of this paper.  

In Folded SUSY, the low energy spectrum below the scale $M$ contains,
in addition to the SM states, the scalar partners of the top quark
which couple just as the stop quarks would in typical SUSY models but
carry no color quantum numbers. The cancellation of the Higgs
mass UV sensitivity is ensured by a supersymmetry broken at the high
scale $M$. The light scalar uncolored top partners, dubbed the
f-stops, realize an accidental supersymmetry of the low energy
spectrum below $M$. Above $M$ the theory contains both colored and
uncolored partners of the SM particles. 

UV completions of Folded SUSY can be obtained in five-dimensional (5D)
supersymmetric theories compactified on an
orbifold~\cite{fldsusy2007}.  Supersymmetry is broken by the
Scherk-Schwarz mechanism~\cite{Scherk:1978ta} by boundary conditions at $M\simeq 1/R$,
where $R$ is the compactification radius. In order to obtain the
accidental SUSY spectrum below $1/R$ with uncolored f-stops instead of
stops, it is necessary to enlarge the gauge symmetry of the SM to
contain at least one extra copy of $SU(3)$. For instance, if
hypermultiplets containing the quark fields are charged under an
$SU(6)$ bulk gauge symmetry, broken by boundary conditions down to
$SU(3)_c\times SU(3)'$, it is possible to arrange the orbifold
procedure in order to obtain fermion zero modes charged under the SM
$SU(3)_c$ and scalar quark zero modes charged under $SU(3)'$.  The f-squarks are charged under the electroweak SM gauge group and couple
to the $W^\pm$, $Z$ and the photon.

The phenomenology of Folded SUSY models presents several novel
features, many of which are shared by other neutral natural models. In
particular, since f-squarks are expected to be  much heavier than
$\Lambda_{\rm QCD'}$ they do not hadronize when pair-produced at
colliders~\cite{Burdman:2008ek}. Instead, they form a highly
excited state that would promptly radiate both photons and $SU(3)'$
glueballs in order to go to the onium ground state, after which it decays
back to SM states.  For instance, bounds on f-squarks can be obtained from LHC data on $W\gamma$
resonances, considering the $\tilde{U}-\bar{\tilde{D}}$ electroweak
pair production resulting in bounds of approximately
$450~$GeV~\cite{Burdman:2014zta}. 
Additionally, the glueballs of $SU(3)'$ might decay
back to SM states through  highly displaced
vertices~\cite{Craig:2015pha,Curtin:2015fna}.   
In this paper we are concerned with the  lepton sector of Folded
SUSY. Although it  plays very little role
in the taming of the Higgs UV sensitivity, the fact that the
f-sleptons are 
typically lighter than the f-squarks and are produced through
similar interactions
makes them  phenomenologically interesting. We will consider minimal
particle content Folded SUSY models and show that the presence of
light f-sleptons is generically unavoidable and even
dangerous. In fact, in order to avoid running into serious trouble
with cosmological bounds, the f-sleptons must decay. This is achieved
through higher dimensional operators, which are naturally present in
the 5D theory. Even then, their lifetimes are typically long on
collider scales leading to interesting bounds. Here we explore the
spectrum and phenomenology of the f-sleptons, the constraints that
they impose on Folded SUSY models and the prospect for new signals. 

The rest of the paper is organized as follows: in the next section we
review the 5D  UV completions of Folded SUSY, in
Section~\ref{sec:fslepton} we discuss the lepton sector and in
Section~\ref{sec:pheno} we study the f-slepton phenomenology at the
LHC. We conclude in Section~\ref{sec:con}.

\section{Folded Supersymmetry from 5D Orbifolds}
 \label{sec:folded}
The aim of a UV completion of Folded SUSY is to have uncolored
stop quarks at low energies that still protect the Higgs mass from
quadratic UV sensitivity. Typically light squarks will be charged under
an $SU(3)$ gauge symmetry other than QCD. This requires extending the
SM gauge sector to accommodate at least two copies of $SU(3)$. 
 In order to obtain this low energy spectrum  we consider a flat
extra dimension compactified on a $S_1/Z_2$ orbifold. In the 5D bulk
the theory is supersymmetric. We assume that Supersymmetry is broken
by boundary conditions by the Scherk-Schwarz
mechanism~\cite{Scherk:1978ta}. 
Since the QCD sector of the 5D theory is extended to include at least
another SU(3), it is possible to carefully choose the boundary
conditions  such that  bulk quark hypermultiplets will have fermion zero modes charged
under QCD, but scalar zero modes charged under the additional $SU(3)$.
We consider here the so-called maximal twist, in which the
 ${\cal N}=1$  supersymmetries   preserved at the extradimensional
 fixed points are related by a maximal $SU(2)_R$ rotation. This choice
 ensures that at any given Kaluza-Klein level fermions and bosons are
 degenerate even if they belong to a different representation of the
 bulk gauge group. This guarantees that the one loop cancellation of the
 top quark contributions to $m_h^2$ is propagated throughout the
 tower.  Although this choice of twist may not have by itself a Higgs
 potential with the correct electroweak vacuum, we will see below that
 this is likely to be solved when considering the presence of
 additional sources of SUSY breaking. Models with less-than-maximal
 twist are considered in Ref.~\cite{Cohen:2015gaa}.

At low energies the couplings of the Higgs to third generation up
quarks and (uncolored) squarks are 
\begin{equation}
(\lambda_t H_uQU +{\rm h.c.} )+ \lambda_t^2 \,|H_u \tilde{Q}|^2 +
\lambda_t^2 |\tilde{U}|^2\,|H_u|^2~.
\label{Hcouplings}
\end{equation}
Then, although SUSY is not present in the low energy spectrum
(e.g. $Q$ and  $\tilde{Q}$ are not related by SUSY transformations),
there is still and accidental supersymmetry that guarantees the
cancellation of quadratic divergences at one loop.

Since SUSY has been broken by boundary conditions, this means that the
part of the spectrum that is absent below the compactification scale
acquires masses of the order of $1/R$.
The Folded SUSY mechanism protects $m_h^2$ at one loop. Two loop
quadratically divergent 
contributions  from the top sector will begin to be of $O(1) \,m_h^2$
at around $5~$TeV. 
Thus, if we choose $1/R\gtrsim 5$~TeV,
the Little Hierarchy problem is solved by the colorless top partners
up until this scale, where the full 5D SUSY spectrum re-appears.
Values of $1/R$ much above this will result in increased fine
tuning. 

The minimal fully realistic model of Folded SUSY has the gauge
symmetry $SU(3)_A\times SU(3)_B\times SU(2)_L\times U(1)_Y\times Z_2$~\cite{Burdman:2008ek}.
Here the $Z_2$ exchanges $A$ and $B$, the $SU(3)_A$ corresponds to the
SM color, whereas $SU(3)_B$ is the mirror color gauge group. The gauge
fields propagate in the bulk of a 5D space compactified on $S_1/Z_2$,
with fixed points at $y=0$ and $y=\pi R$. As mentioned earlier,
SUSY in the 5D bulk results in the following hypermultiplets,
from which the SM quarks and their f-superpartners emerge:
\begin{eqnarray}
\hat{Q}_A = (3,1,2)_{1/6}\quad\quad \hat{Q}_B = (1,3,2)_{1/6}\nonumber \\
\hat{U}_A = (3,1,1)_{-2/3} \quad\quad\hat{U}_B =  (1,3,1)_{-2/3}\nonumber\\
\hat{D}_A = (3,1,1)_{1/3} \quad\quad \hat{D}_B = (1,3,1)_{1/3}
\label{hyperquarks}~,
\end{eqnarray}
where the labels $A$ and $B$ denote SM fields and f-partners
respectively, and we have omitted generation indices.  The quantum
number assignments in (\ref{hyperquarks}) correspond to $SU(3)_A\times
SU(3)_B\times SU(2)_L\times U(1)_Y$. Similarly, the lepton sector is obtained
from the hypermultiplets
\begin{eqnarray}
\hat{L}_A = (1,1,2)_{-1/2}\quad\quad \hat{L}_B = (1,1,2)_{-1/2}\nonumber \\
\hat{E}_A = (1,1,1)_{1} \quad\quad \hat{E}_B = (1,1,1)_{1}
\label{hyperleptons}~.
\end{eqnarray}

The zero-mode folded scalars acquire masses through non-local radiative
corrections in the bulk. The contributions from the gauge interactions
is given by~\cite{Antoniadis:1998sd,Delgado:1998qr}
\begin{eqnarray}
m_{\tilde{Q}}^2 &=& K \frac{1}{4\pi^4}\,\left(\frac{4}{3}g_3^2 +\frac{3}{4}g_2^2
+ \frac{1}{36}g_1^2\right) \frac{1}{R^2}\nonumber\\
m_{\tilde{U}}^2 &=& K \frac{1}{4\pi^4}\,\left(\frac{4}{3}g_3^2 
+ \frac{4}{9}g_1^2\right) \frac{1}{R^2}\nonumber\\
m_{\tilde{D}}^2 &=& K \frac{1}{4\pi^4}\,\left(\frac{4}{3}g_3^2 
+ \frac{1}{9}g_1^2\right) \frac{1}{R^2}\nonumber\\
m_{\tilde{L}}^2 &=& K \frac{1}{4\pi^4}\,\left(\frac{3}{4}g_2^2 
+ \frac{1}{4}g_1^2\right) \frac{1}{R^2}\nonumber\\
m_{\tilde{E}}^2 &=& K \frac{1}{4\pi^4}\, g_1^2\, \frac{1}{R^2}~.
\label{fslpmass}
\end{eqnarray}
There are also contributions coming from Yukawa interactions but these
are only significant for the third generation. For the up-type squarks they result in
\begin{equation}
m_{\tilde{Q}_3}^2 = K \frac{\lambda_t^2}{8\pi^4}\,\frac{1}{R^2}, \qquad\qquad
m_{\tilde{U}_3}^2 = K \frac{\lambda_t^2}{4\pi^4}\,\frac{1}{R^2}~.
\label{fsqrthird}
\end{equation}

In addition to Scherk-Schwarz SUSY Breaking (SSSB), there are likely 
other  sources of SUSY breaking coming from brane-localized Kahler
operators~\cite{Dimopoulos:2014aua,Garcia:2015sfa}. Their origin is
related to the need to dynamically stabilize  the extradimensional
radius with a vanishing four dimensional cosmological constant.  The
desired effect can be parametrized by a singlet brane-localized chiral
superfield $X$ with 
\begin{equation}
F_X\simeq \frac{1}{(\pi R)^2}~,
\label{fx}
\end{equation}
where the F-term is chosen to cancel the typical contributions
to the Casimir energies from bulk fields~\cite{vonGersdorff:2003rq} and contributes additional soft masses. The relevant terms for soft masses
are generically of the form
\begin{equation}
\int d^4\theta
\delta(y)\,\frac{c_\psi}{\Lambda^3}\,X^\dagger X\,\psi^\dagger \psi~,
\label{kahler}
\end{equation}
where $\psi=Q,U,D,L,E$. The coefficients satisfy $c_\psi\sim O(1)$.
 This can result in $O(1)$ corrections to the soft masses in
(\ref{fslpmass}) and (\ref{fsqrthird}), affecting the rather
predictive relations between them and $1/R$. 

The resulting  localized  SUSY-breaking Kahler operators might also be
of importance in order to obtain satisfactory EWSB
through  terms of the form
\begin{equation}
\int d^4\theta
\delta(y)\,\frac{c_{H_i}}{\Lambda^2}\,X^\dagger X\,H_i^\dagger H_i~,
\label{kahlerhiggs}
\end{equation}
with $H_i=H_u, H_d$.

\section{Light Sleptons in Folded SUSY\label{sec:fslepton}}
\begin{figure}[H]
\begin{center}
\includegraphics[scale=0.35]{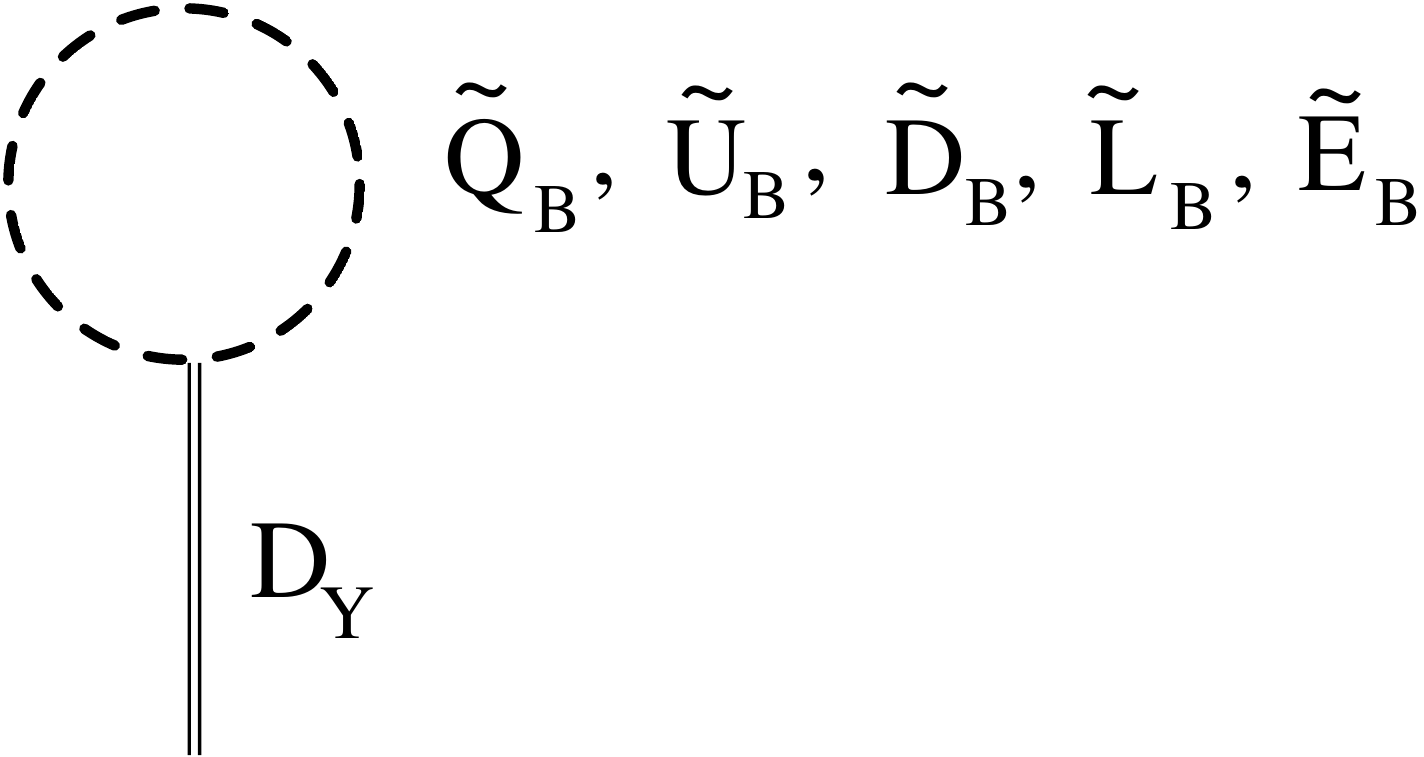}
\caption{
Contributions from $B$-type zero mode scalar fermions to the FI terms of 
hypercharge.  }
\label{fig:filoop} 
\end{center}
\end{figure}

The $A$ and $B$ lepton hypermultiplets in (\ref{hyperleptons}) share
the same gauge quantum numbers. So in principle it appears unnecessary
to introduce this duplication in the spectrum, just having one
of the two as a singlet of the $Z_{AB}$ exchange symmetry would be
enough.  However, this is not possible since it would introduce an 
unwanted Fayet-Iliopoulos (FI) term. The dangerous FI terms come from
the one loop tadpole contributions of scalars to the hypercharge D
term, $D_Y$~\cite{Ghilencea:2001bw} . These are depicted in Figure~\ref{fig:filoop}.
In the absence of $\tilde{L}_B$ and $\tilde{E}_B$ the one-loop FI term
contribution to the Higgs mass squared coming from the B-type squarks is
\begin{equation}
(\delta m_h^2)_{\rm FI} \simeq g'^2\,\frac{3}{16\pi^2}\,\Lambda^2~,
\label{fitomh2}
\end{equation}
where $\Lambda$ is the cutoff of the 5D theory. This
is cancelled by the $\tilde{L}_B$ and $\tilde{E}_B$ contributions. 
For comparison, the minimal model contains a one loop contribution to
$m_h^2$ from the electroweak sector given by
\begin{equation}
(\delta m_h^2)_{\rm gauge} \simeq g'^2\,\frac{3}{16\pi^2}\,\frac{1}{R^2}~,
\label{gaugetomh2}
\end{equation} 
which is cutoff by the gauginos at  $1/R$ instead of
$\Lambda$.  Furthermore, the contribution in (\ref{gaugetomh2}) can be
reduced in extensions of Folded SUSY where the electroweak
sector is embedded in a larger gauge group such as
$SU(4)$~\cite{fldsusy2007}. Thus, the FI terms in (\ref{fitomh2})
would result in a sizable tuning of the Higgs mass which is not really
present even in the minimal model. This can be
avoided by introducing the B hypermultiplet. 
This is a generic feature of Folded SUSY models, as long as FI terms
are to be avoided. 

Given the choice of boundary conditions outlined in
the previous section, the hypermultiplets $\hat{L}_B$ and $\hat{E}_B$
result in scalar zero modes with similar quantum numbers as the MSSM
sleptons. 
As it was pointed out in Ref.~\cite{fldsusy2007}, at the level of
renormalizable interactions  these folded scalar leptons are
stable as long as R-parity is conserved. Also operators  of the type 
$\bar{L}_A H_A E_B + \bar{L}_B H_B E_A $, that could in principle lead
to prompt two-body decays of f-sleptons~\footnote{We thank Z. Chacko
  for pointing this out.}, will not do so in models
where the electroweak sector is also folded such as the $SU(4)$
extension mentioned above.  

 However, non-renormalizable interactions of the type
\begin{equation}
O_1= \delta(y)\int\,d\theta^2\,\left(\frac{Q_AQ_AQ_AL_B}{\Lambda} +
  \frac{Q_BQ_BQ_BL_A}{\Lambda}\right)~,
\label{nonren1}
\end{equation}
and 
\begin{equation}
O_2=\delta(y)\int\,d\theta^2\,\left(\frac{U_AU_AD_AE_B}{\Lambda} +
  \frac{U_BU_BD_BE_A}{\Lambda}\right)~,
\label{nonren2}
\end{equation}
are present and will lead to the decay of the f-sleptons. In
(\ref{nonren1}) and (\ref{nonren2}) $\Lambda$ is the cutoff of the 5D
theory, and flavor indices have been suppressed.
These operators contain a vertex of f-sleptons into two quarks and a
heavy squark all carrying color. Integrating out the heavy squark
results in the decay vertex of f-sleptons into three quarks and the
LSP, which here we are assuming to be the higgsino. In the next section we will study the phenomenology associated
with the decays of f-sleptons.
 
\section{Phenomenology of Folded Sleptons
\label{sec:pheno}}
Written in components the operator $O_1$ in (\ref{nonren1}) is
\begin{equation}
O_1 = \epsilon^{\alpha\beta\gamma}\left(U_i^\alpha D_j^\beta -
  D_i^\alpha U_j^\beta\right)\left(U_k^\gamma E_{B\ell} -D_k^\gamma N_\ell\right)~,
\label{o1incomp}
\end{equation}
where the greek indices denote $SU(3)_c$, the latin indices stand for
flavor, and we are still using superfield notation. For simplicity, we
drop the subindex $A$ when it refers to a SM color-carrying field.   
Similarly, the operator $O_2$ in (\ref{nonren2}) can be expanded into
\begin{equation}
O_2 = \epsilon^{\alpha\beta\gamma}\,U_i^{c\alpha} E_{Bj}^c U_k^{c\beta}
D_\ell^{c\gamma}~.
\label{o2incomp}
\end{equation}
We are interested in extracting the decay vertices for the
f-sleptons. Although in general we could consider the decays of both 
 $\tilde{E}_{BR}$
and $\tilde{N}_B$ (assuming  $\tilde{E}_{BL}$ is heavy enough to
beta-decay to $\tilde{N}_B$)  we will focus on the singlet which is
likely to be much lighter than the doublet, resulting in larger
production cross sections.
As we will see below, the lifetime of
$\tilde{E}_{BR}$ is often long compared to typical detector
sizes. Thus, the most stringent bounds will come from the visible
tracks of $\tilde{E}_{BR}$ going through all or part of the detector. 

Among the vertices that can be extracted from $O_2$ in
(\ref{o2incomp}), we are particularly interested in those containing
two light quarks and a squark. These squarks have
masses of order $1/R$ and when integrated out result in a coupling to
the corresponding quark and the higgsino, which we assume here to be
the LSP.  Thus the effective vertex will be proportional to the Yukawa
coupling of the squark integrated out, and the decay rate should be
dominated by the intermediate stop resulting in a final state with two
jets a top and missing energy whenever kinematically allowed. The
resulting dimension-seven effective operator has the form
\begin{equation}
\frac{c_2\,y_q}{\Lambda\,M_{\tilde{U}_R}^2}\,\tilde{E}_{BR}
\,\epsilon^{\alpha\beta\gamma} q_R^\alpha u_R^\beta
d_R^\gamma\,\tilde{h}^0~,
\label{dim7op}
\end{equation}   
where $c_2\sim O(1)$ and $q_R=u_R, c_R, t_R$ and similarly for $\tilde{U}_R$.

We are now in a position to estimate the lifetime of the right-handed
folded slepton $\tilde{E}_{BR}$. As discussed in Section~\ref{sec:folded}, in the absence of additional sources
of SUSY breaking the f-slepton masses are given by
(\ref{fslpmass}). For instance, for $1/R=5~$TeV we have
$m_{\tilde{E}_{BR}}\simeq 125~$GeV. The dominant decay channel is then
the charm, plus two jets plus missing energy. In order to estimate
the lifetime, we use $\Lambda = 4/R$ and $c_2=1$ to obtain
$\tau_{\tilde{E}_{BR}}\simeq 4\times 10^5~$s. Such a long lifetime has
several problems. Even if we were to ignore the constraints from BBN,
there are direct searches for stable charged particles that put
severe bounds on $m_{\tilde{E}_{BR}}$~\cite{ATLAS:2014fka,Chatrchyan:2013oca}.   
 We derive a mass exclusion from the  CMS constraints on heavy stable
 charged particles.  The production cross section of the
 $\tilde{E}_{BR} $ is the same as the direct electroweak production
 cross section of right handed staus in typical supersymmetric
 scenarios. Assuming stability throughout the detector, the best bound
 is~\cite{Chatrchyan:2013oca} 
\begin{equation}
m_{\tilde{E}_{BR}}>339~{\rm GeV}~ ,
\label{ssbound} 
\end{equation}
at $95\%$~C.L.
This bound applies to the f-slepton since for this value of the mass the
lifetime in this simplified case is $\tau_{\tilde{E}_{BR}} \simeq
10^{-5}~$s, corresponding to $c\tau_{\tilde{E}_{BR}}\simeq 3600~$m. 
The bound is even  more constraining if we take into
account the fact that there are three almost degenerate folded scalar
leptons. The splittings among them come from the Yukawa couplings to
the Higgs and are at most 
\begin{equation}
\delta m^2_{\tilde{E}_B} \simeq K\frac{\lambda^2_\tau}{4\pi^4}\frac{1}{R^2}~,
\label{fstausplit} 
\end{equation}
which results in a f-stau a few GeV heavier than the f-selectron
and the f-smuon for $1/R \sim 5~$TeV. Taking this into account and
multiplying the production cross section by an appropriate factor we
obtain the bound
\begin{equation}
m_{\tilde{E}_{BR}}>420~{\rm GeV}~,
\label{ssboundx3} 
\end{equation}
at $95\%$~C.L.

The bounds in  (\ref{ssbound}) and (\ref{ssboundx3}) are rather stringent  if the spectrum of
folded sfermions is solely determined by Scherk-Schwarz SUSY breaking,
 i.e. by (\ref{fslpmass}) and (\ref{fsqrthird}). For instance, in order to
obtain $m_{\tilde{E}_{BR}}=339~$GeV, we need $1/R\simeq 13~$TeV, which
translates into f-stop masses above $1~$TeV.   For the bound in
(\ref{ssboundx3}) the top partner mass reaches $1.3~$TeV. 
Such large values would reintroduce  significant fine tuning in the
Higgs mass. This situation is relaxed when considering additional
sources of SUSY breaking such as  the ones introduced in
eq.~(\ref{kahler}), since these result typically in order one
corrections to the zero-mode soft masses.
In what follows we will obtain projections for the LHC bounds with
$\sqrt{s}=13$~TeV for both cases, with and without the additional
SUSY-breaking terms.

\begin{figure}[H]
\begin{center}
\includegraphics[scale=0.40]{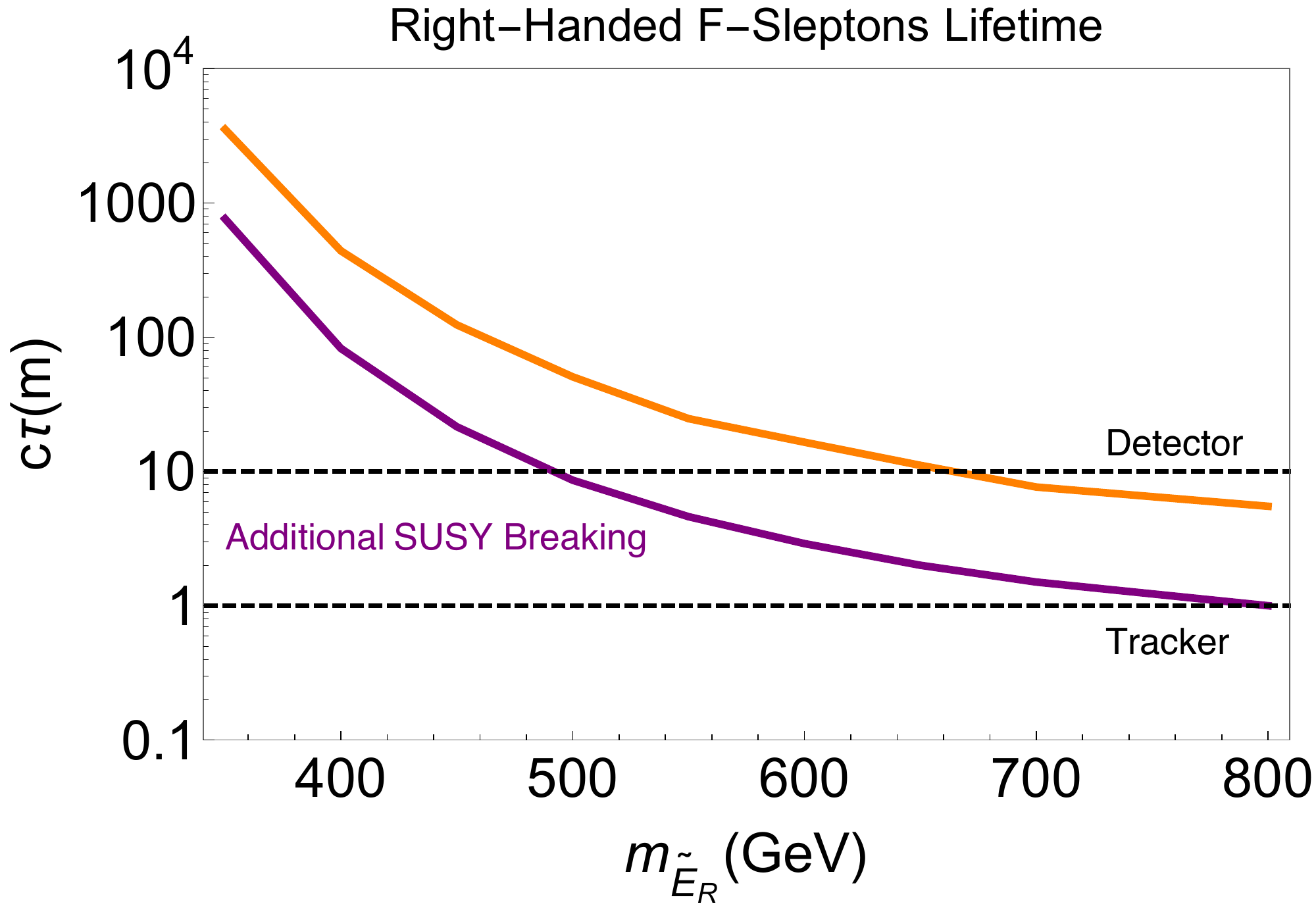}
\caption{Decay length $c\tau_{\tilde{E}_{BR}}$
  vs. the right-handed folded slepton mass. The top curve corresponds
  to the case with the spectrum fully determined by Scherk-Schwarz
  boundary conditions, eq.~(\ref{fslpmass}). The bottom curve is obtained with
additional sources of SUSY breaking localized on the $y=0$ brane. }
\label{fig:ctauvsmass} 
\end{center}
\end{figure}

As mentioned above, the right-handed f-slepton decay in the case
without additional SUSY breaking occurs mostly outside the detector. 
The lifetime of $\tilde{E}_{BR}$ becomes small enough for a significant
fraction of decays to happen inside the detector only for 
$m_{\tilde{E}_{BR}}\gtrsim 650~$GeV. This can be seen from the top curve in
Figure~\ref{fig:ctauvsmass}, where we plot $c\tau_{\tilde{E}_{BR}}$
vs. the right-handed f-slepton mass. 
On the other hand, with  additional SUSY breaking as in eq.~(\ref{kahler}),
the situation changes. To get an idea of the effect we consider
$F_X=1/(\pi R)^2$, and $c_\psi=1$ for all flavors.
As a result a 
significant number of decays occur inside the detector for lighter
masses. This can be seen in the bottom curve of
Figure~(\ref{fig:ctauvsmass}), for masses above $500$~GeV
we have a significant fraction of decays in the outer regions of the detector.   

We can estimate the reach of the heavy stable charged particles search at
$\sqrt{s}=13~$TeV. We extrapolate the bounds obtained by CMS in~\cite{Chatrchyan:2013oca} for $\sqrt{s}=8~$TeV
using a procedure similar to~\cite{CReach}. In our Figures a mass is excluded for a given luminosity if the number of events produced at $\sqrt{s}=13~$TeV is equal or greater than the number needed for the exclusion of stau direct production in the $\sqrt{s}=8~$TeV CMS analysis. In both cases the relevant events are those before cuts, but we count only those for which the f-slepton does not decay inside the detector. The CMS analysis~\cite{Chatrchyan:2013oca} sets a bound also on particles that look stable only inside the tracker. For each luminosity we show the strongest exclusion between the tracker-only and the full detector searches. To determine the distance travelled by the f-sleptons inside the experimental apparatus we use their generator level $\eta$ distribution simulated using {\tt MadGraph 5}~\cite{Alwall:2011uj}. 

\begin{figure}
\begin{center}
\includegraphics[scale=0.40]{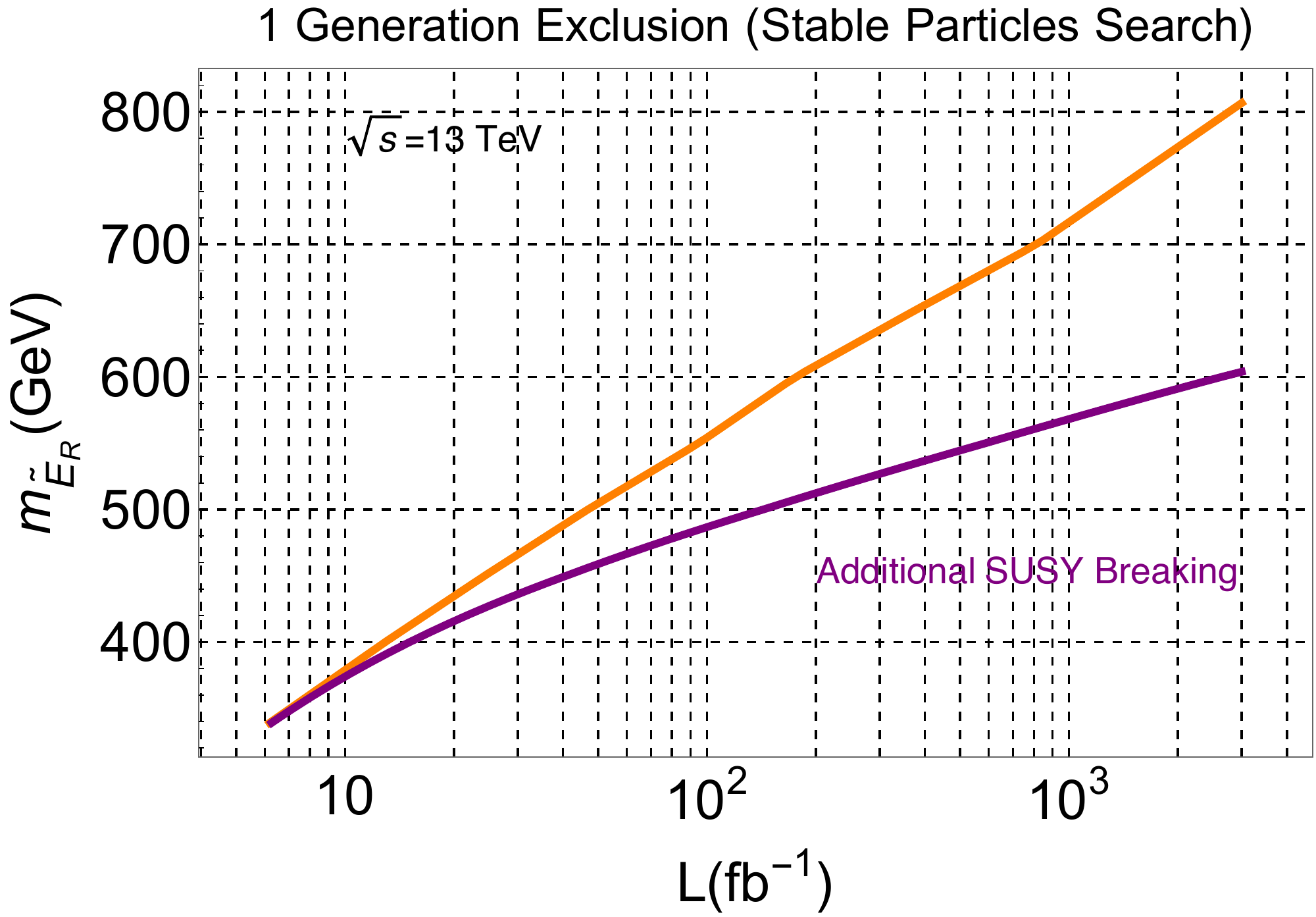}
\caption{Reach of the LHC with $\sqrt{s}=13~$TeV in the f-slepton
  mass, for one f-slepton generation vs. the required integrated luminosity. 
The top curve corresponds to the case with the spectrum fully determined by Scherk-Schwarz
  boundary conditions, eq.~(\ref{fslpmass}). The bottom curve is obtained with
additional sources of SUSY breaking localized on the $y=0$ brane. The
bounds are obtained extrapolating the CMS heavy stable charged particles search~\cite{Chatrchyan:2013oca}.
}
\label{fig:1gen} 
\end{center}
\end{figure}

\begin{figure}
\begin{center}
\includegraphics[scale=0.40]{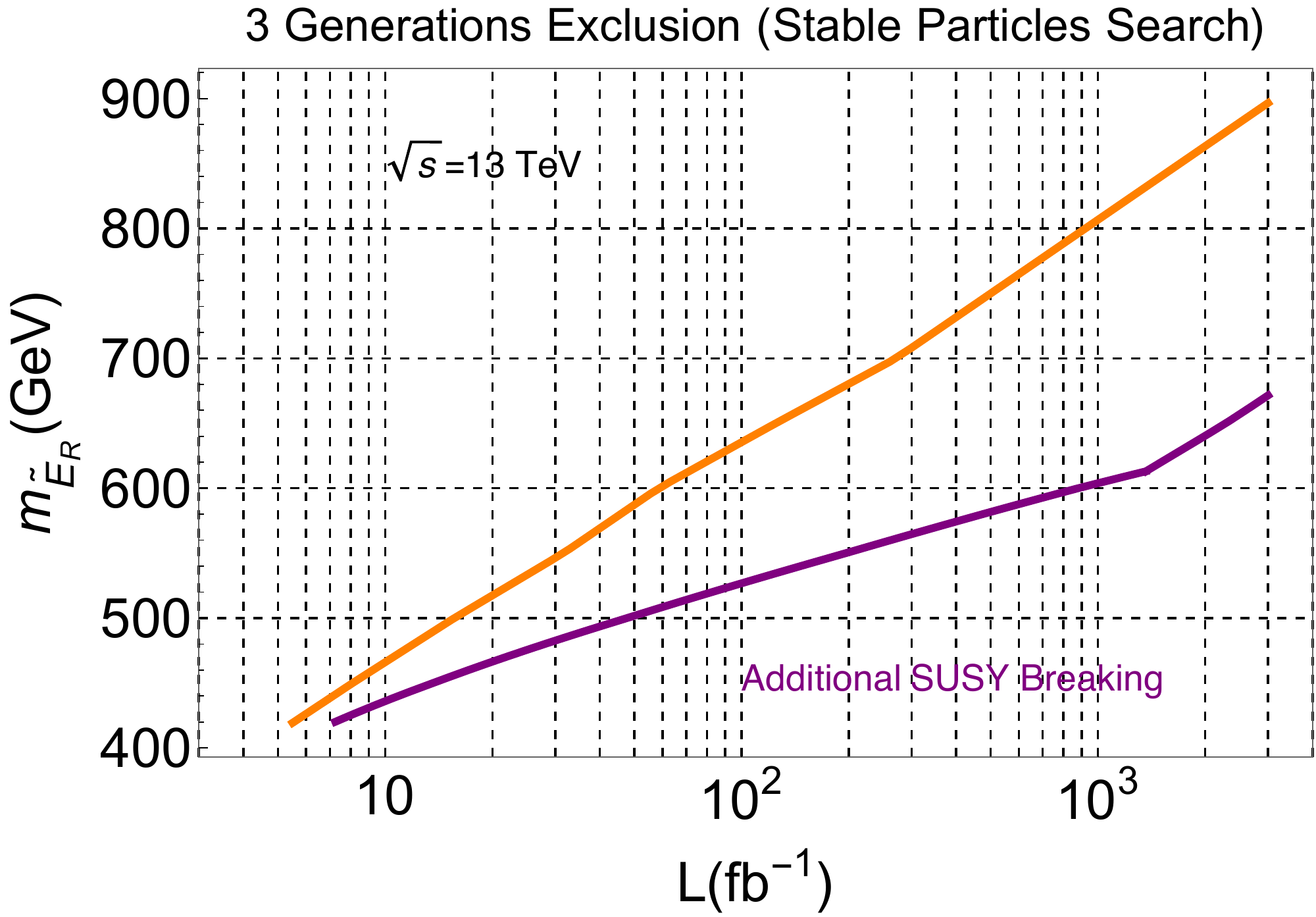}
\caption{Reach of the LHC with $\sqrt{s}=13~$TeV in the f-slepton
  mass, for three degenerate f-slepton generations vs. the required integrated luminosity. 
The top curve corresponds to the case with the spectrum fully determined by Scherk-Schwarz
  boundary conditions, eq.~(\ref{fslpmass}). The bottom curve is obtained with
additional sources of SUSY breaking localized on the $y=0$ brane. The
bounds are obtained extrapolating the CMS heavy stable charged particles search~\cite{Chatrchyan:2013oca}.
}
\label{fig:3gen} 
\end{center}
\end{figure}

In Figure~\ref{fig:1gen} we plot the RH f-slepton mass excluded at $95\%$
C.L. vs the required luminosity for the one generation case. The top curve corresponds to the
original SSSB spectrum, whereas the bottom one includes the additional
sources of SUSY breaking mentioned above. The spectrum fully
determined by SSSB requires $2\;{\rm ab}^{-1}$ to exclude a
$800~$GeV mass. With additional sources of SUSY
breaking a large fraction of events would result in decays inside the
detector. Thus, the stable particles search looses sensitivity and the bounds are degraded, as shown in the bottom
curve of Figure~\ref{fig:1gen}. The tracker-only search has a lower sensitivity and does not improve the bound.
Figure~\ref{fig:3gen} corresponds to the case of three degenerate
species of f-sleptons. However, since the coefficients of additional
SUSY-breaking operators in (\ref{kahler}) may have a non-trivial flavor 
structure~\cite{Garcia:2014lfa} this situation may not be generic. Thus, we should consider
the bottom curve of Figure~\ref{fig:3gen} as an estimate of the  best
case scenario for experiments. Here the tracker-only search for stable particles allows to gain some sensitivity with high integrated luminosity, as signaled by the change in slope of the plot above $1\;{\rm ab}^{-1}$.
If additional supersymmetry breaking is present, the f-sleptons decay
inside the detector most of the time for $m_{\tilde E_{BR}}\gtrsim
500$~GeV and searches for displaced vertices in the inner
tracker, the calorimeters and the muon system become the most
sensitive channels. We have estimated that a tracker-only search
for displaced vertices plus jets as the one described
in~\cite{Aad:2015rba} would exclude slepton masses between $500$ and
$850$~GeV with 3 $\rm{ab}^{-1}$ of integrated luminosity, even if the
peak efficiency on the signal of the kinematical cuts was as low as $5\%$. This shows that the sensitivity lost by the stable particles search can be recovered and the LHC will have roughly the same mass reach in the SSSB case and if additional SUSY breaking is present. Furthermore, in the event of a discovery, including information from the
calorimeters~\cite{Aad:2015asa} and the muon system~\cite{Aad:2015uaa}
could allow a measurement of the slepton lifetime. 
A more detailed estimate of the reach of the LHC for
$\sqrt{s}=13~$TeV, particularly for the case of mostly in-detector
decays is left for future work.  
 
\section{Conclusion}
\label{sec:con}
We have examined the lepton sector of Folded SUSY theories. In
particular, we have shown that it is necessary to include two
sets of lepton hypermultiplets with identical quantum numbers, labeled A and B, in
order to avoid dangerous Fayet-Iliopoulos terms which could
reintroduce a cutoff dependence of the Higgs mass. The resulting
spectrum includes, in addition to the SM leptons, scalar leptons of
the B type, the lightest of which is the hypercharge $Y_{\tilde{ E}_B}=-1$. This f-slepton is stable unless higher dimensional
operators, always present in the UV completions studied here, are
considered. The resulting decay rate is suppressed by the cutoff
$\Lambda$ as well as by the mass of the colored squarks which is of
$O(1/R)$. If SUSY is exclusively broken by the Scherk-Schwarz
mechanism, the relation between the soft masses and the
compactification scale in eq.~(\ref{fslpmass}) results in f-sleptons that are
largely stable on collider scales. From this we can extract a bound on
the f-slepton mass that can be turned into a top-partner mass exclusion. For one
species of f-sleptons this is $m_{\tilde{E}_B}>339~$GeV, which
translates into $M_{\tilde{T}}>1$~TeV. The bound is actually stronger
when considering that there are three nearly degenerate species of
f-sleptons:  $m_{\tilde{E}_B}>420~$GeV and
$M_{\tilde{T}}>1.3$~TeV. This constitutes a stringent constraint on
naturalness. 

The situation can be ameliorated by the presence of additional
SUSY-breaking sources, as shown in eqs.(\ref{kahler}) and
(\ref{kahlerhiggs}). They arise as a consequence of a dynamical
stabilization of the extradimensional radius. 
Their presence may be crucial in obtaining satisfactory EWSB through
terms as in (\ref{kahlerhiggs}).
But centrally for this work, the terms in (\ref{kahler})  introduce additional soft mass terms
with order one uncertainties that break the direct relation between
masses and $1/R$ given by (\ref{fslpmass}). This makes it possible for 
 f-slepton mass to be larger
without increasing correspondingly $1/R$, diluting the power of the
f-slepton mass bounds on the top-partner masses and consequently on naturalness. 
This has the consequence of making f-slepton decays faster, to the
point that decays inside the detectors should be observable during the
$\sqrt{s}=13~$TeV run at the LHC. This work motivates a more careful study of f-slepton decays by the experimental collaborations. However it is possible to estimate that the LHC experiments will be sensitive to masses as large as $(800-850)~$GeV with 3 $\rm{ab}^{-1}$ of integrated luminosity.

{\bf Acknowledgments:}
We thank Z.~Chacko, T. Cohen and D. Pinner for helpful conversations. 
G.B. acknowledges the support
of The Ambrose Monell Foundation as well as the State of S\~{a}o Paulo
Research Foundation (FAPESP). R.T.D. acknowledges support from the Marvin
L. Goldberger Membership and 
DOE grant DE-SC0009988.

\bibliography{fsleptons}

\end{document}